\documentclass[12pt]{amsart}

\usepackage{amsfonts,amsmath,amssymb,amsthm}

\newcommand{\ra}{\rightarrow}
\newcommand{\NN}{{\mathbb N}}
\newcommand{\RR}{{\mathbb R}}
\newcommand{\CC}{{\mathbb C}}
\newcommand{\HH}{{\mathbb H}}

\newcommand{\Symp}{{\mathsf {Symp}}}
\newcommand{\Hilb}{{\mathsf {Hilb}}}
\newcommand{\fdHilb}{{\mathsf {fdHilb}}}
\newcommand{\cS}{{\mathsf S}}
\newcommand{\sone}{{\mathbf 1}}
\newcommand{\Aut}{{\rm Aut}}
\newcommand{\aut}{{\rm aut}}
\newcommand{\ad}{{\rm ad}}
\newcommand{\Ob}{{\rm Ob}}

\newcommand{\Spec}{{\Phi{\rm Spec}}}
\newcommand{\fdQM}{{\sf fdQM}}

\newcommand{\id}{{\rm id}}
\newcommand{\stimes}{\boxtimes}

\newcommand{\tS}{{\circ_S}}
\newcommand{\Mor}{{\rm Mor}}
\newcommand{\sq}{{\rm sq}}

\newtheorem{theorem}{Theorem}[section]
\newtheorem{axiom}{Axiom} 
\newtheorem{prop}{Proposition}[section]

\newtheorem{cor}{Corollary}[section]
\newtheorem{defn}{Definition}[section]
\newtheorem{gerst}{Theorem\!}

\begin{document}

\author{Anton Kapustin}
\address{California Institute of Technology}
\email{kapustin@theory.caltech.edu}
\title{Is there life beyond Quantum Mechanics?}

\begin{abstract} We formulate physically-motivated axioms for a physical theory which for systems with a finite number of degrees of freedom uniquely lead to Quantum Mechanics as the only nontrivial consistent theory. Complex numbers and the existence of the Planck constant common to all systems arise naturally in this approach. The axioms are divided into two groups covering kinematics and basic measurement theory respectively. We show that even if the second group of axioms is dropped, there are no deformations of Quantum Mechanics which preserve the kinematic axioms. Thus any theory going beyond Quantum Mechanics must represent a radical departure from the usual a priori assumptions about the laws of Nature.

\end{abstract}

%\begin{titlepage}

\maketitle

%\end{titlepage}

\section{Introduction}

The axiomatic structure of Quantum Mechanics (QM) has long been a puzzle. Ideally, all mathematical structures and axioms they satisfy should have a clear physical meaning. That is, structures should correspond some natural operations on observables, and axioms should express some natural properties of these operations. Now, if we look at axioms of QM, as formulated for example in \cite{mackey}, the situation is very far from this ideal. The prime offender is axiom VII, which essentially says that observables are bounded Hermitian linear operators in a Hilbert space $V$. What is the physical meaning of the operation of adding two observables? What is the physical meaning, if any, of the associative product of operators on $V$? Why do complex numbers make an appearance, although observables form a vector space over $\RR$? Why should observables be linear operators at all? 

Another way to phrase the question is this. Ideally, axioms should be formulated in such a way that both QM and Classical Mechanics (CM) are particular realizations of these axioms depending on a parameter $\hbar$, and CM can be obtained as a``contraction'' or $\hbar\ra 0$ limit of QM. An axiom like axiom VII of \cite{mackey} is then clearly unacceptable. 

These questions may seem metaphysical rather than physical, akin to ``why is the space three-dimensional?" or "why is there something rather than nothing?". But there is a concrete physical problem where a physically motivated systems of axioms would be very useful. Many people have wondered whether QM is exactly true, or is only an approximation. Accordingly, there have been attempts to construct "nonlinear QM", none of them completely successful even from a purely theoretical standpoint, as far as we know. (For a sample of such attempts see \cite{mielnik, BBM, HB, weinberg, lucke, nattermann}.) One may take the failure of such attempts to indicate that the structure of QM is ``rigid'' and does not admit any physically sensible deformations depending on some ``meta-Planck constant''. But to make this precise and formulate a no-go theorem one first needs to formulate a physically satisfactory set of axioms that any generalization of QM should satisfy. Conversely, a no-go theorem could indicate which physical requirements need to be relaxed when constructing generalizations of QM. 
Another physical issue which such a no-go theorem could clarify is whether it is possible to have a consistent theory which includes both quantum and classical systems. 

In this paper we propose a physically-motivated set of axioms for a physical theory and show that for systems with a finite number of degrees of freedom the only nontrivial possibility is the usual identification of observables with Hermitian operators in Hilbert space. It is also not possible to combine quantum and classical systems in a nontrivial way while preserving the axioms. We also briefly discuss which axioms could be relaxed to allow deviations from QM. We argue that this requires a radical departure from the usual a priori assumptions about physical systems.

There have been numerous works in the past which claim to derive the rules of QM from more basic assumptions \cite{BvonN, Hardy, Fuchs, CAP, muellermasanes, BarnumWilce}. Why propose yet another axiomatization of QM? 
Some of these works suffer from the same problem as axiom VII  in \cite{mackey}, namely they postulate some nontrivial mathematical structure without sufficient physical justification. Others prove ``too much'', in the sense that they derive the standard QM but rule out its useful generalization, QM with superselection rules. Many of these attempts at axiomatization also rule out classical mechanics as a viable theory. 

We will try to explain carefully the physical meaning of every axiom. Our approach is also different from most other approaches in that we focus on the structure of  observables rather than states. In fact, the notion of a state of a physical system does not appear anywhere in our axioms. In the usual QM, such an approach is quite popular and begins by postulating that observables form a $C^*$-algebra. We do not wish to start with such an assumption, because the $C^*$-property is very strong, implying the Uncertainty Principle, and also does  not have a clear physical motivation. In fact, we do not even want to assume that observables form an associative algebra (this again is not well-motivated). Instead, we make two basic physical assumptions: (1)  given two physical systems one can form a composite system; (2) a version of the Noether theorem holds. It was first observed in \cite{GP} that together these two assumptions are quite strong and require the space of observables or its complexification to form an associative algebra. Theorem \ref{thgp} is essentially our interpretation of the results of \cite{GP} in the language of category theory which turns out very convenient for our purposes. In section \ref{measurement} we combine this result with some other natural requirements, like the existence of a spectrum of an observable, and show that if the algebra of observables is finite-dimensional, then it is semi-simple. The well-known Wedderburn theorem then quickly leads one to the conclusion that the only viable possibility is the usual identification of observables with Hermitian operators in a Hilbert space.

In this paper we focus on finite-dimensional systems, but our axioms are designed to apply equally to systems with infinite-dimensional spaces of observables.  While our results are not as strong for such systems, they imply that any physical theory which contains nontrivial finite-dimensional systems must be of the ``quantum'' kind, i.e. the space of observables of every nontrivial system can be identified with the space of Hermitian elements in a non-commutative associative $\star$-algebra. The types of  $\star$-algebras which can occur are  also quite constrained, but we have not attempted to classify them in the infinite-dimensional case.

Our approach largely ignores dynamical issues, focusing on kinematics and measurement theory, and as a consequence has its limitations. In particular we do not discuss states, their time evolution, and the Born rule. Rather, our goal is to explain the fact that observables are Hermitian operators in Hilbert space, and that possible outcomes of measurements are their eigenvalues. A form of the Born rule can then be deduced from the  Gleason's theorem \cite{Gleason}.

\section{A non-technical summary}\label{nontech}

In this section we summarize the results of the paper for the benefit of the reader with an aversion to the language of categories. This will entail some loss of precision.

Our kinematic axioms (section \ref{kinematics}) can be summarized as follows. To each physical system $S$ one can attach a Lie group which describes invertible transformations of variables (in the classical case this is the group of canonical transformations). Let us denote this group $\Aut(S)$ and its Lie algebra $\aut(S)$. Observables are generators of infinitesimal transformations and form a subalgebra of $\aut(S)$. This assures us that to every observable commuting with the Hamiltonian (i.e. to every conserved quantity) one can associate a dynamical symmetry. We regard the connection between symmetries and conserved quantities as one of the fundamental features of both classical and quantum theories, and therefore postulate it in general. Another basic principle is that out of several systems one can form a composite system. The composite of $S_1$ and $S_2$ is denoted $S_1\stimes S_2$. We postulate that observables of individual subsystems commute with each other, and more generally, that from the point of  view of a subsystem $S_1$ observables of the subsystem $S_2$ behave as ordinary numbers. For example, if $A_1$ and $B_1$ are observables for $S_1$, and $C$ is an observable for $S_2$, then the Lie bracket of $A_1\otimes C$ and $B_1\otimes C$ is proportional to $[A_1,B_1]$. From these axioms we deduce that nontrivial physical theories come in three kinds. For a theory of the first kind, the space of observables of every system is a commutative associative algebra, and the Lie bracket is compatible with it in the sense that the Leibniz rule
$$
[f,gh]=[f,g]h+g[f,h]
$$ 
holds for any three observables $f,g,h$. These are classical theories. For a theory of the second kind, the space of observables of every system is an associative algebra over real numbers, and the Lie bracket is proportional to the commutator. The coefficient of proportionality is real and the same for all systems in the theory. One can think of such a theory as a quantum theory with a purely imaginary Planck constant. For a theory of the third kind, the complexification of the space of observables of every system is an associative $\star$-algebra, and the Lie bracket is proportional to the commutator. The coefficient of proportionality is $\sqrt {-1}$ times a real number $1/\hbar$ which is the same for all systems. Quantum Mechanics belongs to this class of theories. 

It is a direct consequence of this result that Quantum Mechanics of finite-dimensional systems cannot be deformed without violating some of the kinematic axioms (section \ref{nogo}).

In section \ref{measurement} we add two more axioms which are designed to enable a sensible interpretation of the theory in macroscopic terms. Namely, we require every observable to have a nonempty spectrum of possible measurement outcomes, and we require every observable with a unique possible measurement outcome to be a constant observable. In the finite-dimensional case these axioms rule out theories of the first and second kind in the above trichotomy.  For finite-dimensional theories of the third kind the axioms force the space of observables to be isomorphic to the space  of Hermitian operators acting on $\CC^n$, or a direct sum of such spaces. The spectrum of an observable is identified with the eigenvalue set of the corresponding Hermitian operator. Thus if nontrivial finite-dimensional systems exist, then both the emergence of complex numbers and Quantum Mechanics are an inevitable consequence of our axioms. 

\section{Axioms: kinematics}\label{kinematics}

{\defn A (physical) theory is a groupoid $\cS$ (i.e. a category all of whose morphisms are invertible) with some additional structures and properties described in the axioms below.  Objects of $\cS$ are called (physical) systems, morphisms of $\cS$ are called kinematic equivalences.}

Commentary: Kinematic equivalences are essentially changes of variables describing a physical system. Thus separation between kinematics and dynamics is implicit in this definition. In the case of CM, the category of physical systems is the category of symplectic manifolds $\Symp$, with symplectomorphisms as kinematic equivalences. In the case of QM, the category of physical systems is the category of Hilbert spaces $\Hilb$, with unitary isomorphisms as kinematic equivalences. A subcategory of the latter category is the category $\fdHilb$, whose objects are finite-dimensional Hilbert spaces. We will call the corresponding theory \fdQM. Note that one need not assume that all unitary isomorphisms are allowed in the case of $\Hilb$ or \fdQM. This takes into account the possibility of superselection sectors. In the case of CM, this possibility is implicitly taken into account by allowing disconnected symplectic manifolds. 

{\axiom (Smoothness). For any two physical systems $S_1,S_2\in \Ob(\cS)$ the set $\Mor(S_1,S_2)$ is a smooth manifold (possibly infinite-dimensional), and the composition of morphisms is a smooth map. }

Commentary: 

(1) Loosely speaking, this means that kinematic equivalences may depend on continuous parameters. This is the case both in CM and QM, and it is rather natural to assume this in general. A better justification is that without such an axiom one can neither formulate continuous dynamics (Axiom 2) nor define observables (Axiom 5) so that a version of the Noether theorem holds. 

(2) There are several versions of the notion of an infinite-dimensional Lie group, such as a Banach Lie group or a Fr\'{e}chet Lie group. Since we will mostly deal with the case when the group of automorphisms $\Aut(S)=\Mor(S,S)$ is finite-dimensional, we will not specify which version we use. 

{\axiom (Continuous dynamics). Time is continuous and parameterized by points of $\RR$. Time evolution of a system $S\in \Ob(\cS)$ is a homomorphism of Lie groups $\RR\ra \Aut(S)$ whose generator is called the Hamiltonian.}

Commentary: This axiom is optional as it is never used in what follows. However, we will use the notion of a Hamiltonian to motivate other axioms. 

{\axiom  (Composite systems). The category $\cS$ is given a symmetric monoidal structure with a tensor product denoted $\stimes$. The identity object is called the trivial system and is denoted $\sone$.  The homomorphism $\Aut(S_1)\times \Aut(S_2)\ra \Aut(S_1\stimes S_2)$ which is part of the monoidal structure is injective.}

Commentary: 

(1) The product $S_1\stimes S_2$ of systems $S_1$ and $S_2$ is called the composite of $S_1$ and $S_2$.  In the case of CM, the product is the Cartesian product of phase spaces. In the case of QM, it is the tensor product of Hilbert spaces. The assumption of having a symmetric monoidal structure implies in particular that we can consider several copies of the same system, i.e. we can form composites $S\stimes \ldots \stimes S$, and that permutation group acts on such composites by automorphisms. 

(2) The trivial system is a physical system which has a unique state. Combining it with any other system $S$ gives a system isomorphic to $S$.

(3) The definition of a monoidal structure on a category implies that for any $S_1,S_2\in \Ob(\cS)$ we are given a homomorphism of Lie groups $\Aut(S_1)\times \Aut(S_2)\ra \Aut(S_1\stimes S_2)$. That is, changes of variables for individual systems give rise to changes of variables for their composite. The injectivity condition says that a nontrivial change of variable for individual systems is a nontrivial change of variables for the composite. 

(4) We will assume that systems are distinguishable. This does not mean that we cannot incorporate systems with identical particles (bosons or fermions) into our framework. Rather, this means that we will not regard a system of $N$ indistinguishable particles as a composite of $N$ one-particle systems. This is especially natural if we regard indistinguishable particles are excitations of a quantum field; then one should regard the field itself, rather than its one-particle excitations, as a separate physical system.

We will denote by $\aut(S)$ the Lie algebra of the Lie group $\Aut(S)$. 

{\axiom (Observables). The set of observables $O(S)$ of a physical system $S$ is a Lie sub-algebra of $\aut(S)$. This sub-algebra is invariant under the adjoint action of $\Aut(S)$.}

Commentary:

(1) We would like to identify an observable with a physical apparatus which measures it. But both in CM and QM an observable is also a dynamical variable, i.e. it can be used to deform the Hamiltonian. In fact, a measuring process is usually modeled by a composite system consisting of a physical system $S$ and a measuring apparatus $M$, such that the Hamiltonian of $S\stimes M$ contains a term proportional to the observable $A\in O(S)$ that one is measuring.  According to Axiom 2, the Hamiltonian is an element of the Lie algebra $\aut(S)$, so a physical observable is also an element of this Lie algebra. We do not assume that all deformations correspond to physical observables. However, since the set of observables must be invariant under automorphisms of the system, $O(S)$ must be a Lie sub-algebra of $\aut(S)$, and in fact an ideal in $\aut(S)$. 

(2) In the case of CM, $O(S)$ is the Lie algebra of Hamiltonian vector fields on the phase space $S$. In the case of QM and \fdQM, $O(S)$ is the Lie algebra of (bounded) anti-Hermitian operators with respect to the commutator. 

(3) This axiom ensures that a version of the Noether theorem holds. Namely every observable in $O(S)$ is a generator of a continuous kinematic symmetry, and an observable which is preserved by the time evolution generates a dynamical continuous symmetry (i.e. a one-parameter subgroup of $\Aut(S)$ which commutes with the time evolution).

{\axiom (Constant observables). For each $S\in \Ob(\cS)$ we are given a distinguished nonzero element $\id_S\in O(S)$ which lies in the center of the Lie algebra $O(S)$. Observables of the form $\lambda\cdot \id_S$ , $\lambda\in\RR$, are called constant observables. We have $O(\sone)=\RR$, with the distinguished element being $1\in\RR$. }

Commentary: A constant observable $\lambda\cdot \id_S$ corresponds to a measuring device whose output is always $\lambda$, regardless of the state of the system $S$. 
The trivial system has a unique state, therefore its only observables are constant observables. Note that $\Aut(\sone)$ is necessarily a commutative Lie group (since $\sone$ is the identity object in a monoidal category), and this axiom implies that its dimension is at least $1$.

{\axiom (Linearization). We are given a functor from $\cS$ to the category of real vector spaces  which for all $S\in \Ob(\cS)$ maps $S$ to $O(S)$. This functor is compatible with the symmetric  monoidal structure, and in particular for every $S_1,S_2\in \Ob(\cS)$ we have a linear map $p_{S_1,S_2}: O(S_1)\otimes O(S_2)\ra O(S_1\stimes S_2)$. This map is required to be injective. Furthermore, $p_{S_1,S_2}(\id_{S_1} \otimes A)$ is the image of $A\in O(S_2)$ under the homomorphism of Lie algebras $\aut(S_2)\ra \aut(S_1\stimes S_2)$. }

Commentary: 

(1) For any $S\in \Ob( \cS)$ the space of observables $O(S)$ is a real vector space. Here and below $\otimes$ denotes tensor product over $\RR$.

(2) We will sometimes shorten $p_{S_1,S_2}$ to $p_{12}$.

(3) There should clearly be a map $p_{12}: O(S_1)\times O(S_2)\ra O(S_1\stimes S_2)$. This map assigns to $(A_1,A_2)\in O(S_1)\times O(S_2)$  the observable obtained by multiplying the outputs of devices measuring $A_1$ and $A_2$. The reason $p_{12}$ should be bilinear is a bit more complicated. First of all, we assume that the output of a device measuring $\lambda A$ is $\lambda$ times the result of measuring $A$. Therefore $p_{12}(\lambda A_1,A_2)=p_{12}(A_1,\lambda A_2)$. Consider now an apparatus which can measure the observable $\lambda A_1$ for any $\lambda\in \RR$. It has a classical lever which controls the choice of $\lambda$. If simultaneously we measure an observable $A_2\in O(S_2)$, we can use the result of the measurement to set the lever position.  Alternatively, we can set the lever position to $1$ and multiply the result of measuring $A_1$ by the result of measuring $A_2$. It is very natural to assume that this gives the same result. That is, measuring observables of $S_2$ does not affect the observables of $S_1$, and the results of measuring the former behave as ordinary numbers as far as $S_1$ is concerned. 
Thus as far the system $S_1$ is concerned, the observable $A_1\otimes A_2$ can be thought as $A_1$ rescaled by a scalar $\lambda$ (the result of measuring $A_2$), and since for any $A_1,B_1\in O(S_1)$ and any scalar $\lambda$ we have an identity
$$
\lambda(A_1+B_1)=\lambda A_1+\lambda B_1,
$$
we must similarly identify $p_{12}(A_1+B_1,A_2)$ and $p_{12}(A_1,A_2)+p_{12}(B_1,A_2)$. This implies that the map $p_{12}$ is bilinear and therefore gives rise to a linear map from $O(S_1)\otimes O(S_2)$ to $O(S_1\stimes S_2)$.  It is also natural to require this map to be injective (if the product of $A_1$ and $A_2$ gives zero, regardless of the state of the systems $S_1$ and $S_2$, then at least one of the observables $A_1,A_2$ must be zero, and therefore $A_1\otimes A_2=0$). Other properties of the maps $p_{12}$ which are hidden in the statement that it comes from a symmetric monoidal functor come from its interpretation as the multiplication map and the associativity and commutativity of ordinary multiplication.

(4) The last requirement arises as follows. Consider a one-parameter subgroup of $\Aut(S_2)$ whose generator is some observable $A\in O(S_2)$. One can think of $A$ as a particular deformation of the Hamiltonian of $S_2$. Given any $S_1\in \Ob(\cS)$, we have the corresponding one-parameter family in $\Aut(S_1\stimes S_2)$ which we can think of as evolution generated by a deformation of the Hamiltonian of the composite system. Clearly, this deformation is simply $A$ but regarded as an observable of the composite system, i.e. $p_{12}(\id_{S_1}\otimes A)$.

(5) Examples such as $\Symp$ and $\Hilb$ show that in general $p_{12}$ is not an isomorphism. However, further axioms will ensure that the image of $p_{12}$ is a Lie sub algebra of $O(S_1\stimes S_2)$. 

(6) Axioms 1-6 imply that linearity over $\RR$ is built into the structure of physical observables. This linearity of observables is different from the linearity (over $\CC$) of states postulated by the superposition principle of QM. Rather, it arises from an identification of observables with infinitesimal deformations of the Hamiltonian.  Our goal is to show that the Hilbert space structure can be deduced from the linearity of observables and some additional natural axioms.

(7) Weinberg's nonlinear QM \cite{weinberg} does not satisfy Axiom 6. Indeed, in Weinberg's nonlinear QM physical systems correspond to Hilbert spaces, and $O(S)_\CC$ consists of homogeneous degree-1 functions on the Hilbert space. Weinberg assumes that the Hilbert space of the composite system is the tensor product of individual Hilbert spaces, as in the usual QM. But there is no reasonable way to define a product of two homogeneity-1 functions on Hilbert spaces $V_1$ and $V_2$ to get a homogeneity-1 function on $V_1\otimes_\CC V_2$. For this reason Weinberg only works with ``additive'' observables which are sums of observables for individual systems. This is clearly unsatisfactory.

{\axiom  For any $S_1,S_2\in \Ob(\cS)$ there exists a function $\sq_{S_1,S_2}: O(S_2)\ra O(S_2)$ such that for any $A_1,B_1\in O(S_1)$ and $C\in O(S_2)$  one has $[p_{12}(A_1\otimes C),p_{12} (B_1\otimes C)]=p_{12}([A_1,B_1]\otimes \sq_{S_1,S_2}(C))$.}

Commentary:

(1) The function $\sq_{S_1,S_2}: O(S_2)\ra O(S_2)$ is defined uniquely if $O(S_1)$ is a nonabelian Lie algebra. If $O(S_1)$ is abelian, then $\sq_{S_1,S_2}$ is arbitrary. 

(2) This axiom is a reflection of the same principle that was used to justify the existence of the map $p_{12}$: from the point of view of system $S_1$, observables of system $S_2$ behave as ordinary numbers. In particular, the observables $p_{12}(A_1\otimes C)$ and $p_{12}(B_1\otimes C)$ can be thought of as $A_1$ and $B_1$ rescaled by the result of measuring $C$, and their Lie bracket should be $[A_1,B_1]$ rescaled by the result of measuring the square of $C$, i.e. $p_{12}$ applied to the product of $[A_1,B_1]$ and the square of $C$. Thus $\sq_{S_1,S_2}: O(S_2)\ra O(S_2)$ should be interpreted as an operation of squaring an observable in $O(S_2)$. In fact, it would be natural to require it to depend only on $S_2$, not $S_1$, but we will see below that this follows automatically from the associativity of $\stimes$, provided there exist systems with a nonabelian Lie algebra of observables.

From now on we will focus on physical theories satisfying Axioms 1-7.  

{\prop \label{two} Let $\cS$ be a theory satisfying Axioms 1-7. For any $S,S'\in\Ob(\cS)$ the image of $p_{S,S'}$ is a Lie sub-algebra of $O(S\stimes S')$.  There exist maps $\tau^{(1)}_{S,S'}: O(S)\otimes O(S)\ra O(S)$ and $\tau^{(2)}_{S,S'}: O(S')\otimes O(S')\ra O(S')$ such that $\forall A, B \in O(S)$ and $\forall A', B'\in O(S')$ we have
\begin{equation}\label{liebracket}
p_{S,S'}^{-1}([p_{S,S'}(A\otimes A'), p_{S,S'}(B\otimes B')])=[A, B]\otimes \tau^{(1)}_{S,S'} (A',B')+\tau^{(2)}_{S,S'}(A, B)\otimes [A', B'].
\end{equation}
The map $\tau^{(1)}_{S,S'}$ (resp. $\tau^{(2)}_{S,S'}$)  is unique if $O(S')$ (reps. $O(S)$) is nonabelian and arbitrary otherwise. In the case when it is unique, it is symmetric, equivariant with respect to  $\Aut(S)$ (resp. $\Aut(S')$), and satisfies the normalization condition $\tau^{(1)}_{S,S'}(A,\id_S)=A$ for any $A\in O(S)$ (resp. $\tau^{(2)}_{S,S'}(A',\id_{S'})=A'$ for any $A'\in O(S')$). Whenever they are well-defined, the maps satisfy $\tau^{(1)}_{S,S'}=\tau^{(2)}_{S',S}$.
}

\begin{proof} 

Consider the expression
$$[p_{S,S'}(A\otimes A'), p_{S,S'}(B\otimes B')].$$
It is a quadrilinear function of $A,B,A',B'$ which is skew-symmetric with respect to the exchange of $A,A'$ and $B,B'$. We can write uniquely it as a sum of two quadrilinear functions, the first one symmetric in $A,B$ and skew-symmetric in $A',B'$, the second one skew-symmetric in $A,B$ and symmetric in $A',B'$:
$$
[p_{S,S'}(A\otimes A'), p_{S,S'}(B\otimes B')]=f^{+-}(A,B;A',B')+f^{-+}(A,B;A',B').
$$
The function $f^{-+}$ is determined by its values on the partial diagonal $A'=B'$. Using Axiom 7, we get
$$
f^{-+}(A,B;C,C)=p_{S,S'}([A,B]\otimes \sq_{S,S'}(A')),\quad \forall C\in O(S').
$$

Similarly
$$
f^{+-}(D,D;A',B')=p_{S,S'}(\sq_{S',S}(D)\otimes [A',B']),\quad \forall D\in O(S).
$$
Hence both $f^{+-}$ and $f^{-+}$ take values in the image of $p_{S,S'}$, and thus the image of $p_{S,S'}$ is a Lie sub-algebra of $O(S\stimes S')$. Furthermore, we see that the Lie bracket on this Lie sub-algebra is given by Eq.~(\ref{liebracket}) with 
\begin{align*}
\tau_{S,S'}^{(1)}(A',B') &=\frac12(\sq_{S,S'}(A'+B')-\sq_{S,S'}(A')-\sq_{S,S'}(B')),\\
 \tau_{S,S'}^{(2)}(A,B) &=\frac12(\sq_{S',S}(A+B)-\sq_{S',S}(A)-\sq_{S',S}(B)).
\end{align*}
Note that this implies that the functions $\sq_{S,S'}$ and $\sq_{S',S}$ are quadratic.

The equivariance of $\tau^{(1)}$ and $\tau^{(2)}$ with respect to the action of $\Aut(S)$ and $\Aut(S')$ is a consequence of the fact that $p_{S,S'}$ is part of the data defining a symmetric monoidal functor. The relation $\tau^{(1)}_{S,S'}=\tau^{(2)}_{S',S}$ is also obvious. To deduce the normalization condition, let $B=\id_S$. Since $\id_S$ is in the center of the Lie algebra $O(S)$, we have
\begin{equation}\label{nnn}
[p_{S,S'}(A\otimes A'), p_{S,S'}(\id_S\otimes B')]=p_{S,S'}(\tau^{(1)}_{S,S'}(A,\id_S)\otimes [A',B']).
\end{equation}
Now we note that according to Axiom 6, $p_{S,S'}(\id_S\otimes B')$ is the generator of a one-parameter subgroup of $\Aut(S\stimes S')$ which acts on $O(S\stimes S')$ by the automorphisms of $S'$ via the adjoint representation. Thus for any $t\in\RR$ we have
$$
\exp(t\ p_{S,S'}(\id_S\otimes B')) p_{S,S'}(A\otimes A') \exp(-t\ p_{S,S'}(\id_S \otimes B'))=p_{S,S'}(A \otimes \exp(t B') A' \exp (-t B')),
$$
which implies 
$$
[p_{S,S'}(A\otimes A'), p_{S,S'}(\id_S\otimes B')]=p_{S,S'}(A \otimes [A',B']).
$$
Assuming that $S'$ is nonabelian we can choose $A',B'$ so that $[A',B']$ is nonzero. Comparing with Eq.~(\ref{nnn}) , we conclude that $\tau^{(1)}_{S,S'}(A,\id_S)=A$. Exchanging $S$ and $S'$ we also get $\tau^{(2)}_{S,S'}(A',\id_{S'})=A'$ for all $A'\in O(S')$ provided $S$ is nonabelian.

\end{proof}

Systems with an abelian Lie algebra of observables are not very interesting since their dynamics is trivial thanks to Axiom 2. 

{\defn A physical theory $\cS$ is called trivial if for all $S\in \Ob(\cS)$ the Lie algebra $O(S)$ is abelian. Otherwise it is called nontrivial.}

In what follows we will focus on nontrivial theories.

{\prop \label{three} Let $\cS$ be a nontrivial theory satisfying axioms 1-7. The map $\tau^{(1)}_{S,S'}$ is independent of $S'$ provided $S'$ is nonabelian. The map $\tau^{(2)}_{S,S'}$ is independent of $S$ provided $S$ is nonabelian. 
}

\begin{proof}

Since for any $S$ and $S'$ the map $p_{S,S'}$ identifies $O(S)\otimes O(S')$ with a Lie sub-algebra of $O(S\stimes S')$, we get a Lie algebra structure on $O(S)\otimes O(S')$ whose explicit form is given by Prop. \ref{two}. For any three systems $U, V, W\in \Ob(\cS)$ starting with a given Lie algebra structure on $O(U\stimes V\stimes W)$ we can therefore define two Lie algebra structures on $O(U)\otimes O(V)\otimes O(W)$, corresponding to two different ways of placing parentheses: $(O(U)\otimes O(V))\otimes O(W)$ vs. $O(U)\otimes (O(V)\otimes O(W))$. These two Lie algebras structures must coincide.  Indeed, the functor which sends $S$ to $O(S)$ is monoidal, therefore we must have 
$$
p_{U\stimes V, W}(p_{U,V}(u\otimes v)\otimes w)=p_{U, V\stimes W}(u\otimes p_{V,W}(v\otimes w)).
$$

Consider therefore any three systems $U,V,W\in \Ob(\cS)$ and any $u,u'\in O(U), v,v'\in O(V), w,w'\in O(W)$. Computing the commutator
$$
[u\otimes v\otimes w,u'\otimes v'\otimes w']
$$
in two different ways, symmetrizing in $w,w'$ and anti-symmetrizing in $v,v'$, we get:
$$
\tau^{(1)}_{U,V\otimes W}(u,u')\otimes [v,v']\otimes \tau^{(2)}_{U,W}(w,w')=\tau^{(1)}_{U,V}(u,u')\otimes [v,v']\otimes \tau^{(2)}_{U\otimes V, W}(w,w').
$$
Let us choose $V$ so that $O(V)$ is nonabelian and choose $v,v'$ so that $[v,v']\neq 0$. Then we get
$$
\tau^{(1)}_{U,V\otimes W}(u,u')\otimes \tau^{(2)}_{U,W}(w,w')=\tau^{(1)}_{U,V}(u,u')\otimes \tau^{(2)}_{U\otimes V, W}(w,w').
$$
It is easy to see that $O(U)\otimes O(V)$ as well as $O(V)\otimes O(W)$ are nonabelian as well (by Axiom 6, they contain Lie sub-algebras isomorphic to $O(V)$). Suppose $O(U)$ is also nonabelian. Then all the maps in the above equation are well-defined. Letting  $u=u'=\id_U$ and using the normalization condition for $\tau^{(1)}_{U,V}$ from Proposition \ref{two}, we get
$$
\tau^{(2)}_{U,W}=\tau^{(2)}_{U\otimes V,W}.
$$
This equality holds for arbitrary $W$ and arbitrary nonabelian $U$ and $V$. Therefore
$$
\tau^{(2)}_{U,W}=\tau^{(2)}_{V,W}
$$
for arbitrary nonabelian $U$ and $V$. Thus $\tau^{(2)}_{U,W}$ does not depend on $U$ provided $O(U)$ is nonabelian. Exchanging $U$ and $W$, we get that
$\tau^{(1)}_{W,U}$ does not depend on $U$ provided $O(U)$ is nonabelian.

\end{proof}

Since $\tau^{(1)}_{W,U}$ does not depend on $U$, from now on we denote it simply by $\tau_W$. Then $\tau^{(2)}_{U,W}=\tau^{(1)}_{W,U}=\tau_W$. Thus each $O(S)$ is equipped with a symmetric bilinear operation $\tau_S:O(S)\otimes O(S)\ra O(S)$, and the Lie algebra structure on $O(S_1)\otimes O(S_2)$ is determined by the Lie algebra structure on $O(S_1)$ and $O(S_2)$, as well as the operations $\tau_{S_1}$ and $\tau_{S_2}$. To decrease the notational clutter,  we will sometimes denote $\tau_S(A,B)$ by $A\tS B$. We also define
$$
(A,B,C)_S=(A\tS B)\tS C-A\tS (B\tS C),
$$
this is the associator for the product $\tS$. 

{\prop \label{four} Let $\cS$ be a nontrivial theory satisfying axioms 1-7, and let $S$ be an arbitrary system in $\cS$. For any $A\in O(S)$ the map $\ad_A: O(S)\ra O(S), B\mapsto [A,B],$ is a derivation of the bilinear operation $\tau_S$, i.e. for any $A,B,C\in O(S)$ we have
$$
[A,B\tS C]=[A,B] \tS C+B\tS [A,C].
$$
}

\begin{proof} Let $g(t)=\exp(tA)$ be the one-parameter subgroup of $\Aut(S)$ generated by $A$. $\Aut(S)$-invariance of $\tau_S$ implies $\tau_S(Ad_{g(t)}B,Ad_g(t)C)=Ad_{g(t)}\tau_S(B,C)$. Differentiating with respect to $t$ and setting $t=0$ gives the desired result. \end{proof}

Propositions \ref{two}, \ref{three} and \ref{four} mean that for any nontrivial theory$\cS$ the collection of triples $(O(S),[\ ,\ ],\tS)$, $S\in \Ob(\cS)$ forms what is called in \cite{GP} " a composition class of two-product algebras". Therefore we can use the results of \cite{GP}.

{\prop  Let $\cS$ be a nontrivial theory satisfying axioms 1-7. For any $S\in\Ob(\cS)$ there exists a pair of numbers $\lambda_S,\mu_S\in\RR$ not simultaneously equal to zero and defined up to an overall scaling such that for all $A,B,C\in O(S)$ we have
\begin{equation}\label{associator}
\lambda_S (A,B,C)_S =\mu_S [[A, C],B].
\end{equation}
For any $S,S'\in \Ob(\cS)$ we have $(\lambda_S: \mu_S)= (\lambda_{S'}:\mu_{S'})$.
}
\begin{proof}
For completeness, we give the proof from \cite{GP}. Let $S,S'\in\Ob(\cS)$. Imposing the Jacobi identity for the Lie bracket on $O(S)\otimes O(S')$ and using the Jacobi identity for the Lie bracket on $O(S),O(S')$ and the fact that $\ad_A.\ad_{A'}$ are derivations of $\tau_S$ and $\tau_{S'}$, respectively, we get
$$
(A\tS B)\tS C\otimes [[A', B'], C']+[[A, B], C]\otimes (A'\tS B')\tS C'+{cycl}=0.
$$
Here $A,B,C\in O(S)$ and $A',B',C'\in O(S')$ are arbitrary elements, and $cycl$ means simultaneous cycling permutations of letters $A,B,C$ and $A',B',C'$. Symmetrizing with respect to $A$ and $B$ we get
\begin{multline}
\left((A,B,C)_S+(B,A,C)_S\right)\otimes [[A',B'],C']= \\ \left([[B,C],A]+[[A,C],B]\right)\otimes (A',C',B')_{S'}.
\end{multline}
Therefore there exist $\lambda_{S'},\mu_{S'}$ not equal to zero simultaneously and defined up to an overall scale such that
\begin{align*}
& \lambda_{S'} (A',C',B')_{S'} =  \mu_{S'} [[A',B'],C'] ,&\\
& \lambda_{S'}\left((A,B,C)_S+(B,A,C)_S\right) = \mu_{S'} \left([[B,C],A]+[[A,C],B]\right).&
\end{align*}

Exchanging $S$ and $S'$ we get the same equations with $A,B,C$ exchanged with $A',B',C'$ and $\lambda_{S'},\mu_{S'}$ replaced with $\lambda_S,\mu_S$. Hence $(\lambda_S: \mu_S)= (\lambda_{S'}:\mu_{S'})$.

\end{proof}

Following \cite{GP}, we can use this result to classify the types of physical theories that can occur.

{\theorem  \label{thgp} Let $\cS$ be a nontrivial theory satisfying axioms 1-7. The following three-fold alternative holds:

(1) For any $S\in\Ob(\cS)$ $\tau_S$ defines a commutative associative product on $O(S)$. Thus $O(S)$ is a commutative Poisson algebra over $\RR$.

(2) There exists $\hbar\in\RR_+$ such that for all $S\in \Ob(\cS)$ the bilinear operation $(A,B)\mapsto A\tS B+\hbar [A,B]$ defines an associative product on $O(S)$. Thus $O(S)$ is an associative algebra over $\RR$.

(3) There exists $\hbar\in\RR_+$ such that for all $S\in\Ob(\cS)$ the bilinear operation  $(A,B)\mapsto A\tS B+i \hbar [A,B]$ defines an associative product on 
$O(S)_\CC=O(S)\otimes\CC$. Thus $O(S)_\CC$ is an associative algebra over $\CC$. 

In the cases (1) and (2) (resp. case (3)) the algebra $O(S)$ (resp. $O(S)_\CC$) is unital for all $S$, with $\id_S$ being the unit element.}

\begin{proof}
If $(\lambda_S:\mu_S)=(0:1)$ for all $S$, then for all $S$ and all $A,B,C\in O(S)$ we have $[[A,B],C]=0$. Let us apply this to the composite of two systems $S$ and $S'$. For arbitrary $A,B\in  O(S)$ and $A', B'\in O(S')$ we compute
$$
0=[[A\otimes A', B\otimes \id_{S'}],\id_S\otimes B']=[A,B]\otimes [A',B'].
$$
If we choose $S'$ to be nonabelian, this means that $S$ is abelian, and vice versa. This means that all $S$ are abelian, which contradicts the fact that $\cS$ is nontrivial. Thus one cannot have $(\lambda_S:\mu_S)=(0:1)$. 

Since $\lambda_S=0$ is impossible,  we can set $\lambda_S=1$ for all $S$ by a rescaling. Then there are three case: $\mu_S=0$, $\mu_S<0$ and $\mu_S>0$. They correspond to the cases (1), (2), and (3). Indeed, if $\mu_S=0$ for all $S$, then $(A,B,C)_S=0$ for all $S$, and the triple $(O(S),\tS, [\ ,\ ])$ is a commutative Poisson algebra. If $\mu_S<0$, we let $\hbar=\sqrt {-\mu_S}$ and define $A\cdot B=A\tS B+\hbar [A,B]$. Using the Jacobi identity, the derivation property and Eq. (\ref{associator}), one can easily check that the dot-product is associative. If $\mu_S>0$, we let $\hbar=\sqrt \mu_S$ and define $A\cdot B=A\tS B+i\hbar [A,B]$. Then the dot-product is again associative. It is obvious that $\id_S$ is the identity element for $O(S)$ or $O(S)_\CC$.  

\end{proof}

This important theorem shows that either $O(S)$ or $O(S)_\CC$ is an associative algebra, something which looks rather mysterious otherwise. Furthermore, in the case (3) $O(S)_\CC$ is a $\star$-algebra, i.e. there is an anti-linear involution $O(S)_\CC\ra O(S)_\CC,$ $A\mapsto A^\star$ such that $(A\cdot B)^\star=B^\star \cdot A^\star$ for all $A, B\in O(S)_\CC$. The involution is given by $(A_1+i A_2)^*=A_1-i A_2$, where $A_1, A_2\in O(S)$.  This theorem also shows that it is impossible to combine  classical systems with a nontrivial dynamics (which correspond to case (1) in the above theorem) and quantum systems (which correspond to case (3)) within a single theory satisfying Axioms 1-7. 

In the next section we will see that once some additional axioms are imposed and the existence of finite-dimensional systems is assumed, cases (1) and (2) become impossible.  In the case (3) the same axioms imply that $O(S)_\CC$ must be isomorphic to a sum of matrix algebras over $\CC$, which means that we are dealing with the usual QM. Thus our approach also explains the origin of complex numbers in QM.

\section{Axioms: measurements}\label{measurement}

Both in CM and QM observables are both dynamical variables and measurables. That is, (1) for any dynamical variable one can find a measuring device producing a real number output, and (2) given any such measuring device, there is a dynamical variable corresponding to it. The former requirement is a part of any Copenhagen-like interpretation of the theory. The meaning of the latter requirement is less obvious and requires some comment. Given a measuring device we can feed its output into a classical computer and get another measuring device.  If the computer is running a program which computes a real function of a single real variable, this means that given any $f:\RR\ra\RR$ and any $A\in O(S)$ we can define a new observable $f(A)\in O(S)$ so that $(f\circ g)(A)=f(g(A))$. We also must have $f(\lambda \cdot \id_S)=f(\lambda)\cdot \id_S$. In what follows it will be sufficient to consider polynomial functions of observables.

{ \defn Let $V$ be a vector space over $\RR$ with a distinguished nonzero element $e\in V$. A polynomial calculus on $V$ is a collection of maps $K_f: V\ra V$ for each polynomial function $\RR\ra\RR$ such that

(1) $K_f(K_g(v))=K_{f\circ g}(v)$ for all polynomial functions $f,g$ and all $v\in V$,

(2) $K_f(\lambda e)=f(\lambda) e$ for all polynomial functions $f$ and all $\lambda\in \RR$,

(3) $K_{f+g}(v)=K_f(v)+K_g(v)$ for all polynomial functions $f,g$ and all $v\in V$,

(4) If $f(x)=\lambda x$ for some $\lambda\in\RR$, then $K_f(v)=\lambda v$ for all $v\in V$.}

For the reasons explained above, we expect that on every $O(S)$ there is a polynomial calculus equivariant with respect to $\Aut(S)$, with $\id_S$ being the distinguished element. Such a calculus is uniquely determined by the squaring operation, i.e. by $K_{x^2}$. Indeed, once one knows how to define arbitrary linear and quadratic functions of elements of $V$, one can recursively define higher powers using the identity
$$
x^{n+1}=\frac12 ((x+x^n)^2-x^2-(x^n)^2).
$$
On the other hand, as explained in the commentary to Axiom 7, the squaring operation is given by $A\mapsto \tau_S(A)$. Hence the polynomial calculus on $O(S)$ is completely determined. Equivalently, this is the polynomial calculus arising from the associative algebra structure on $O(S)$ or $O(S)_\CC$ (because $\tau_S(A)$ is the square of $A$ with respect to the associative product on $O(S)$ or $O(S)_\CC$). We want this polynomial calculus to have reasonable properties compatible with its physical interpretation. This motivates two more axioms.

{\axiom (Physical spectrum of an observable). For any observable $A\in O(S)$ we are given a nonempty  subset of $\RR$ called the physical spectrum of $A$ and denoted $\Spec (A)$. For any polynomial function $f:\RR\ra \RR$ one has $\Spec ( f(A))=f(\Spec(A))$. The physical spectrum of a constant observable $\lambda\cdot \id_S$ is the one-point set $\{\lambda\}$. }

Commentary: $\Spec(A)$ is the set of possible results of measuring an observable $A$. Clearly it must be nonempty. Measuring a constant observable $\lambda\cdot  \id_S$ always gives $\lambda$. 

{\axiom (No phantom observables).  Let $A\in O(S)$. If $\Spec(A)=\{\lambda\}$ for some $\lambda\in\RR$, then $A=\lambda\cdot \id_S$. }

Commentary: If measuring an observable always gives the same result, regardless of the state of the system, then such an observable must be a constant observable. 

These two axioms put strong constraints on $O(S)$ if $O(S)$ is finite-dimensional. Since nontrivial systems with a finite-dimensional space of observables exist in nature (spin systems), it is important to analyze this case. Let us say that a theory $\cS$ is finite-dimensional if $O(S)$ is finite-dimensional for all $S\in \Ob(\cS)$. We would like to classifying nontrivial finite-dimensional theories. We will now show that the only such theory is \fdQM.

{\prop \label{roots} Let $\cS$ be a nontrivial theory satisfying Axioms 1-8, let $S\in \Ob(\cS)$, and let $A\in O(S)$ satisfy $P(A)=0$, where $P:\RR\ra \RR$ is a polynomial function. Then $\Spec(A)$ is a subset of the real roots of $P(x)$. In particular, the set of real roots of $P$ is nonempty. }
\begin{proof}
$P$ maps $\Spec(A)$ to the point $0$,  therefore $\Spec(A)$ is contained in the zero set of $P$. 
\end{proof}

{\cor\label{nilpotent} Let $\cS$ be a nontrivial theory satisfying Axioms 1-9. If $A\in \Ob(\cS)$ satisfies $A^n=0$ for some  $n\in\NN$, then $A=0$.}
\begin{proof}
Immediate consequence of Prop. \ref{roots} and Axiom 9.
\end{proof}

{\cor Let $\cS$ be a nontrivial theory satisfying Axioms 1-8, and  let $S\in \Ob(\cS)$ be a system such that $O(S)$ is finite-dimensional. The physical spectrum of any $A\in O(S)$ is a finite nonempty subset of $\RR$ whose cardinality does not exceed $\dim O(S)$. }
\begin{proof}
Since $O(S)$ is finite-dimensional, for a sufficiently large $N$ not exceeding $\dim O(S)$ the set of observables $1,A,A^2,\ldots,A^N$ will be linearly dependent. Thus there exists a polynomial function $P:\RR\ra\RR$  of degree less or equal than $\dim O(S)$ such that $P(A)=0$.
\end{proof}

Corollary \ref{nilpotent} can be used to rule out cases (1) and (2) of Theorem \ref{thgp} if $\cS$ is a nontrivial finite-dimensional theory. Indeed, we have the following well-known fact from algebra.

{\theorem If $V$ is a finite-dimensional algebra over $\RR$ with no nonzero nilpotent elements, then $V$ is isomorphic to a sum of several copies of $\RR$, $\CC$ or $\HH$, where $\HH$ is the quaternion algebra. If $V$ is in addition commutative, then $V$ is isomorphic to a sum of several copies of $\RR$ and $\CC$. }

\begin{proof} Since $V$ is finite-dimensional, its Jacobson ideal consists of nilpotent elements (see e.g. \cite{Cohn}, Theorem 5.3.5) and thus is trivial. Hence $V$ is semi-simple, and by the Wedderburn theorem \cite{Cohn} is a direct sum of matrix algebras over $\RR$, $\CC$ or $\HH$. But a matrix algebra over a ring can be free of nilpotents only if it is the ring itself. 
\end{proof}

{\cor (The inevitability of complex numbers). Let $\cS$ be a nontrivial finite-dimensional theory satisfying Axioms 1-9. Then cases (1) and (2) of Theorem \ref{thgp} are impossible.}
\begin{proof}
Suppose $\cS$ belongs to cases (1) or (2). Thus for every $S\in\Ob(\cS)$ the algebra $O(S)$ is free of nilpotents, therefore it is a finite sum of several copies of $\RR$, $\CC$ and $\HH$. But $O(S)$ cannot contain summands isomorphic to $\CC$ or $\HH$ because they have elements $A$ which satisfy $A^2+1=0$,  which would contradict Proposition \ref{roots}. Thus in both case (1) and case (2) $O(S)$ is isomorphic to a sum of several copies of $\RR$. In the case (2) this immediately implies that the Lie bracket on $O(S)$ vanishes. In the case (1) the Lie bracket on $O(S)$ also vanishes because for any $A\in O(S)$  $ad_A$ must be a derivation, and the only derivation of $\RR\oplus\ldots\oplus \RR$ is zero.  This contradicts the assumption that $\cS$ is nontrivial. 
\end{proof}

To deal with the case (3) of Theorem \ref{thgp} let us classify finite-dimensional $\star$-algebras over $\CC$ with no nonzero nilpotent Hermitian elements. A matrix algebra $M_n(\CC)$ with $\star$ given by the usual Hermitian conjugation (conjugate transpose), $v\mapsto v^\dagger$, is an example of such a $\star$-algebra. Another example is $\CC\oplus\CC$ with the $\star$ given by
$$
\star: (a,b)\mapsto (b^*,a^*).
$$
Its Hermitian elements have the form $(a,a^*)$ where $a\in\CC$ is arbitrary. Let us call this $\star$-algebra $V_2$. The following theorem shows that these are essentially the only examples. 

{\theorem If $V$ is a finite-dimensional $\star$-algebra over $\CC$ with no nonzero nilpotent Hermitian elements, then $V$ is isomorphic to a sum of several copies of matrix algebras over $\CC$, with the standard $\star$-structure, and several copies of $V_2$.}

\begin{proof}
First let us show that $V$ is semi-simple. Let $v$ belong to the Jacobson radical of $V$. This means that $1-a v b$ has a two-sided inverse for all $a,b\in V$. Therefore $v^*$ is also in the Jacobson radical, and so are $v+v^*$ and $i(v-v^*)$. But all elements in the Jacobson radical are nilpotent \cite{Cohn}, hence we must have $v=0$. Thus $V$ is a semi-simple algebra, and by the Wedderburn theorem is isomorphic to a direct sum of matrix algebras over $\CC$.

It remains to classify allowed $\star$-structure on $V$. Any two $\star$-structures on $V$ differ by an algebra automorphism of $V$.  
An automorphism of a semi-simple algebra can be decomposed into a composition of a permutation of isomorphic simple summands and automorphisms of individual summands. Thus it is sufficient to consider the case when $V$ is a direct sum of $k$ copies of $M_n(\CC)$. In this case the most general $\star$-structure must have the form
\begin{multline*}
v=(v_1,\ldots,v_k)\mapsto v^*=(m_1 v_{P(1)}^\dagger m_1^{-1},\ldots, m_k v_{P(k)}^\dagger m_k^{-1}), \\
 v_1,\ldots,v_k \in M_n(\CC),
\end{multline*}
where $m_1,\ldots,m_k$ are invertible elements of $M_n(\CC)$ and $P$ is a permutation of the set $\{1,\ldots,k\}$. Requiring the square of this transformation to be the identity transformation shows that $P$ can contain only cycles of length $1$ and $2$. Also, if $P$ can be decomposed as a product of $N$ disjoint cycles of lengths $k_1,\ldots,k_N$, $k_1+\ldots+k_N=k$, then clearly the $\star$-algebra $V$ decomposes as a direct sum of $N$ $\star$-algebras, each of which is a sum of several copies of $M_n(\CC)$, with $\star$ cyclically permuting the summands. Combining these two observations, we see that it is sufficient to consider two cases: the case when $k=2$, $V=M_n(\CC)\oplus M_n(\CC)$, and the $\star$ permuting the two summands, and the case when $V=M_n(\CC)$. 

In the former case, the $\star$-operator acts by
$$
(v_1,v_2)\mapsto (H^{-1} v_2^\dagger H, H^{-1} v_1^\dagger  H),
$$
where $H\in M_n(\CC)$ is invertible and satisfies $H=H^\dagger$. Hermitian elements in such an algebra have the form $(a, H^{-1} a^\dagger H)$, where $a\in M_n(\CC)$ is arbitrary.  If $n>1$, this algebra has nonzero Hermitian nilpotent elements (just take $a$ to be a nonzero nilpotent matrix). Thus we must have $n=1$, which means that $V$ is isomorphic to $V_2$. 

In the latter case $V=M_n(\CC)$ and the $\star$-structure has the form
$$
v\mapsto H^{-1} v^\dagger H,
$$
where $H$ is invertible and satisfies $H=H^\dagger$. If we think of $M_n(\CC)$ as the algebra of linear endomorphisms of $\CC^n$, then $v^\star$ is the adjoint of $v$ with respect to a sesquilinear form on $\CC^n$
$$
\langle x, y\rangle=x^\dagger Hy,\quad x,y\in\CC^n. 
$$
Note that $H$ and $\lambda H$  give rise to identical $\star$-structures for any nonzero $\lambda\in \RR$. The isomorphism class of this $\star$-structure is determined by the absolute value of the signature of $H$.  Thus we may assume that $H$ is diagonal with eigenvalues $\pm 1$. If it has two eigenvalues with opposite signs, then $V$ contains a $\star$-sub-algebra isomorphic to $M_2(\CC)$ with the $\star$-structure
$$
\begin{pmatrix} a & b \\ c & d\end{pmatrix}\mapsto \begin{pmatrix} a^* & -c^* \\ -b^* & d^*\end{pmatrix},\quad a,b,c,d\in\CC.
$$
The latter algebra has a nonzero nilpotent Hermitian element
$$
\begin{pmatrix} 1 & 1 \\ -1 & -1\end{pmatrix}.
$$
Hence the eigenvalues of $H$ must all have the same sign, which means that the $\star$-structure on $V$ is isomorphic to the standard one.  

Therefore any finite-dimensional $\star$-algebra $V$  with no nonzero nilpotent Hermitian elements is a direct sum of matrix algebras over $\CC$ with the standard $\star$-structure and several copies of $V_2$. 
\end{proof}

The $\star$-algebra $V_2$ cannot occur as a summand of $O(S)_\CC$. Indeed, $V_2$ contains a Hermitian element $A=(i,-i)$ satisfying $A^2+1=0$, which would contradict Proposition \ref{roots}. Thus we get

{\cor (The inevitability of Quantum Mechanics). Let $\cS$ be a nontrivial finite-dimensional theory satisfying Axioms 1-9. Then for all $S\in\Ob(\cS)$ $O(S)_\CC$ is isomorphic as a $\star$-algebra to a direct sum of matrix algebras over $\CC$, with the the standard $\star$-structure. This isomorphism identifies $O(S)$ with the subspace of Hermitian matrices, and the Lie bracket on $O(S)$ is mapped to $-i/\hbar$ times the commutator, where $\hbar$ is the same for all $S\in\Ob(\cS)$. The physical spectrum of an observable $A\in O(S)$ is the set of its eigenvalues.}
\begin{proof}
The only thing which needs to be proved is that $\Spec(A)$ is the set of all eigenvalues of $A$ (since $A$ is a Hermitian operator, the set of its eigenvalues is nonempty and real), rather than some proper subset. Recall that if $\{\lambda_1,\ldots,\lambda_K\}\subset \RR$ is the set of eigenvalues of a Hermitian operator $A$,  then $A$ satisfies the equation $P(A)=0$, where $P(x)$ is a real polynomial with simple roots $\lambda_1,\ldots,\lambda_K$, and there is no polynomial function of lower degree which annihilates $A$. On the other hand, if, say, $\lambda_1$ were not in $\Spec(A)$, then the polynomial function 
$$
f(x)=\prod_{i=2}^K (x-\lambda_i)
$$
would map $\Spec(A)$ to zero, and therefore by Axioms 8 and 9 we would have $f(A)=0$. This is impossible, since $f$ has degree $K-1$.
\end{proof}

Thus Axioms 1-9 imply that observables in any nontrivial finite-dimensional theory are described by Hermitian operators on a Hilbert space, perhaps with some superselection rules imposed, and the physical spectrum of an observable is the set of its eigenvalues. The group $\Aut(S)$ contains all unitary transformations compatible with superselection rules.

\section{A no-go theorem for nonlinear QM}\label{nogo}

In this section we ask how one can relax the above axioms to avoid the conclusion that QM is inevitable. For example, could one drop Axiom 9? That is, could there be dynamical variables which are trivial as far as measurements are concerned (measuring them always gives zero), but are nonzero elements of $O(S)$? Such ``phantom'' observables could provide a novel kind of ``hidden variables'''. Even more radically, one could question the assumption that an arbitrary polynomial function of an observable is again an observable and drop Axioms 8 and 9 altogether (although we would probably want to retain some notion of the spectrum of an observable).

Nevertheless even then one can prove an interesting no-go theorem if one asks a more modest question. Namely, could there be small corrections to the rules of QM depending on a small parameter? This is a much easier question to deal with because there is a well-known theorem \cite{MG}:

{\gerst  A finite-dimensional semi-simple algebra over a field is rigid (does not admit nontrivial infinitesimal deformations).}

Thus we can immediately conclude that in any deformation of the theory \fdQM\ satisfying Axioms 1-7 only, the algebras $O(S)_\CC$, $S\in\Ob(\cS)$, would still be isomorphic to a sum of matrix algebras over $\CC$. The space of observables $O(S)$ is then the space of Hermitian elements in this algebra with respect to a $\star$-structure. The operation $\tau_S$ and the Lie bracket are given by the anti-commutator and $-i/\hbar$ times the commutator, respectively. Thus to classify deformations of the two-product algebra $O(S)$ it is sufficient to classify deformations of the $\star$-structure on a direct sum of matrix algebras over $\CC$.

{\prop Let $V$ be a direct sum of matrix algebras over $\CC$. The standard $\star$-structure on $V$ given by $v\mapsto v^\dagger$ does not admit nontrivial  infinitesimal deformations.}

\begin{proof}
Any two $\star$-structures on $V$ differ by an automorphism of $V$. If they are infinitesimally close, then the corresponding automorphism is infinitesimally close to the identity element. It is easy to see that such an automorphism must act on each simple summand separately, therefore it is sufficient to consider the case $V=M_n(\CC)$. Any automorphism of $M_n(\CC)$ is inner, so the most general $\star$-structure on $M_n(\CC)$ is given by $v\mapsto H^{-1} v^\dagger H$, where $H\in V$ is Hermitian and invertible. In other words, the $\star$-structure is given by the adjoint with respect to a sesquilinear form on $\CC^n$
$$
\langle x, y\rangle=x^\dagger H y.
$$
The isomorphism class of such a $\star$-structure is determined by the absolute value of the signature of $H$. If $H$ is infinitesimally close to $1$, then it is positive-definite, and therefore its signature is $n$, i.e. the same as for the standard $\star$-structure. Hence any $\star$-structure on $V$ infinitesimally close to the standard one is isomorphic to it.
\end{proof}

{\cor  (No-Go for Nonlinear QM). Finite-dimensional Quantum Mechanics does not admit nontrivial infinitesimal deformations within the class of theories satisfying Axioms 1-7.}

What conclusions can we draw from all this for the prospects of constructing a nonlinear deformation of Quantum Mechanics? One important assumption was that dynamical variables form a Lie algebra. This was motivated by the desire to have a traditional formulation of the Noether theorem. One could try to find some weakening of this axiom so that the Noether theorem only holds in some limit. Another assumption was that given several systems $S_1, S_2,\ldots, S_N$ one can form a composite system $S_1\stimes \ldots \stimes S_N$. Perhaps this is approximately true for systems small compared to the size of the Universe, but fails for large systems.  Most conservatively, one might try to turn to systems with an infinite number of degrees of freedom,  i.e. systems with an infinite-dimensional space of dynamical variables $O(S)$. But even then our results show that all finite-dimensional systems are described by Quantum Mechanics exactly. 
Thus a theory which goes beyond QM must violate at least one of Axioms 1-7 and therefore represent a radical departure from the usual a-priori assumptions about the structure of physical laws. 

\section*{acknowledgments}

This work was supported in part by the Department of Energy grant DE-FG02-92ER40701.


\begin{thebibliography}{99}


\bibitem{mackey} G.~W.~Mackey, ``Mathematical Foundations of Quantum Mechanics,'' Dover Books on Physics (2004). 

\bibitem{mielnik} B.~Mielnik, ``Generalized quantum mechanics,'' Commun.\ Math.\ Phys.\  {\bf 37}, 221 (1974).

\bibitem{BBM} I.~Bialynicki-Birula and J.~Mycielski, ``Nonlinear Wave Mechanics,'' Annals Phys.\  {\bf 100}, 62 (1976).

\bibitem{HB} R.~Haag and U.~Bannier, ``Comments on Mielnik's generalized (non linear) quantum mechanics'', Comm.\ Math.\ Phys.\ {\bf 60}, 1 (1978).

\bibitem{weinberg}  S.~Weinberg, ``Testing Quantum Mechanics,''  Annals Phys.\  {\bf 194}, 336 (1989).

\bibitem{lucke} W.~Lucke, ``Nonlinear Schrodinger dynamics and nonlinear observables,'' quant-ph/9505022.

\bibitem{nattermann} P.~Nattermann, ``Generalized quantum mechanics and nonlinear gauge transformations,''
  quant-ph/9703017.
  
  \bibitem{BvonN} G.~ Birkhoff, J.~von~Neumann, ``The Logic of Quantum Mechanics,'' Ann. Math. {\bf 37}, 823 (1936). 

\bibitem{Hardy} L.~Hardy, ``Quantum theory from five reasonable axioms,'' quant-ph/0101012.

\bibitem{Fuchs} C.~A.~Fuchs, ``Quantum Mechanics as Quantum Information (and only a little more),'' quant-ph/0205039.

\bibitem{CAP} G.~Chiribella, G.~M.~D'Ariano, P.~Perinotti, ``Probabilistic theories with purification,'' Phys. Rev. {\bf A81}, 062348 (2010) [arXiv:0908.1583[quant-ph]].

\bibitem{muellermasanes} L.~Masanes and M.~P.~Muller, ``A derivation of quantum theory from physical requirements,'' 
  New J.\ Phys.\  {\bf 13}, 063001 (2011)
  [arXiv:1004.1483 [quant-ph]].
  
\bibitem{BarnumWilce} H.~Barnum, A.~Wilce, ``Local tomography and the Jordan structure of quantum theory,'' arXiv:1202.4513 [quant-ph].


\bibitem{GP} E.~Grgin and A.~Petersen, ``Algebraic Implications of Composability of Physical Systems,''
  Commun.\ Math.\ Phys.\  {\bf 50}, 177 (1976).
  
\bibitem{Gleason} A. M.~Gleason, ``Measures on the closed subspaces of a Hilbert space,'' J.\ Math.\ Mech. {\bf 6}, 885 (1957).
  
\bibitem{Cohn} P.~M.~Cohn, ``Basic Algebra: Groups, Rings, and Fields,'' Springer (2005). 

\bibitem{MG} M.~Gerstenhaber, ``On the deformation of rings and algebras,'' Ann.\ Math.\ (2)\ {\bf 79}, 59 (1964).

\end{thebibliography}
\end{document}